\documentclass[aps, prl,10pt,twocolumn,superscriptaddress,nofootinbib]{revtex4-1}
\usepackage{mathrsfs, amssymb, amsmath}  
\usepackage{epsfig, cancel}
\usepackage{latexsym}
\usepackage{natbib, comment}
\usepackage{url}
\usepackage{dcolumn}
\usepackage{multirow}
\usepackage{color}
\usepackage{cancel}
\usepackage{soul}
\usepackage[normalem]{ulem}
\usepackage{amsfonts,amssymb,amsmath, txfonts}
\usepackage{graphicx,epsfig}
\usepackage{psfrag}
\usepackage{hyperref}
\hypersetup{colorlinks=true}
\usepackage{mathtools}
\usepackage{enumitem}
\usepackage{float}
\usepackage[dvipsnames]{xcolor}
\usepackage{xcolor}
\hypersetup{ linktoc=all,
    colorlinks, linkcolor={blue},
    citecolor={darkred}, urlcolor={darkgreen}
}
\definecolor{rosy}{RGB}{230,235,252}
\definecolor{myframetitle}{RGB}{90,89,170}
\definecolor{myblocktitle}{RGB}{140,185,249}
\definecolor{mytitle}{RGB}{10,80,26}

\definecolor{darkgreen}{RGB}{27,130,45}
\definecolor{darkblue}{rgb}{0,0,0.3}
\definecolor{darkred}{rgb}{0.7,0,0}

\definecolor{light gray}{RGB}{220,220,220}
\definecolor{dark purple}{RGB}{108,0,217}
\definecolor{pink}{RGB}{190,20,100}
\definecolor{orang}{RGB}{193,63,0}
\definecolor{green}{RGB}{11,98,17}
\definecolor{darkpink}{RGB}{153,0,76}
\definecolor{bluegreen}{RGB}{0,102,102}
\definecolor{greenlagan}{RGB}{0,102,0}
\definecolor{redgreen}{RGB}{102,102,0}
\definecolor{Redgreen}{RGB}{153,76,0}
\definecolor{vividviolet}{rgb}{0.62, 0.0, 1.0}
\definecolor{amaranth}{rgb}{0.9, 0.17, 0.31}
\definecolor{palatinateblue}{rgb}{0.15, 0.23, 0.89}
\definecolor{brightpink}{rgb}{1.0, 0.0, 0.5}
\definecolor{cornflowerblue}{rgb}{0.39, 0.58, 0.93}
\definecolor{deepcarminepink}{rgb}{0.94, 0.19, 0.22}
\definecolor{radicalred}{rgb}{1.0, 0.21, 0.37}

\def\H0{{\text{H}\hspace*{-2.05mm}\text{H} 0\hspace*{-1.35mm}0\ }}

\def\be{\begin{equation}}
\def\ee{\end{equation}}
\def\beq{\begin{equation}}
\def\eeq{\end{equation}}
\def\bea{\begin{eqnarray}}
\def\eea{\end{eqnarray}}
\newcommand{\dd}{\textrm{d}}

\begin{document}

\title{On the Analysis Dependence of DESI Dynamical Dark Energy}

\author{Eoin \'O Colg\'ain}
\affiliation{Atlantic Technological University, Ash Lane, Sligo, Ireland}
\author{Saeed Pourojaghi} 
\affiliation{School of Physics, Institute for Research in Fundamental Sciences (IPM), P.O.Box 19395-5531, Tehran, Iran}
\author{M. M. Sheikh-Jabbari} 
\affiliation{School of Physics, Institute for Research in Fundamental Sciences (IPM), P.O.Box 19395-5531, Tehran, Iran}

\begin{abstract}
We continue scientific scrutiny of the DESI dynamical dark energy (DE) claim by explicitly demonstrating that the result depends on the analysis pipeline. Concretely, we define a likelihood that converts the $w_0 w_a$CDM model back into the (flat) $\Lambda$CDM model, which we fit to DESI constraints on the $\Lambda$CDM model from DR1 Full-Shape (FS) modelling and BAO. We further incorporate CMB constraints. Throughout, we find that $w_0$ and $w_a$ are within $1 \sigma$ of the $\Lambda$CDM model. Our work makes it explicit that, in contrast to DR1 and DR2 BAO, there is no dynamical DE signal in FS modelling, even when combined with BAO and CMB. Moreover, one confirms late-time accelerated expansion today $(q_0 < 0)$ at $ \gtrsim 3.4 \sigma$ in FS modelling + BAO. On the contrary, DR1 and DR2 BAO fail to confirm $q_0 < 0$ under similar assumptions. Our analysis highlights the fact that trustable scientific results should be independent of the analysis pipeline.  
\end{abstract}

\maketitle

\section{Introduction}
The DESI collaboration has presented a suite of dynamical DE claims \cite{DESI:2024mwx, DESI:2024hhd, DESI:2025zgx}. These claims begin with baryon acoustic oscillation (BAO) data, but become more statistically significant in the presence of external datasets. Overlooked in all this discussion is the observation from \cite{DESI:2024jis} that DESI Full-Shape (FS) modelling of galaxy clustering \textit{exhibits no hint of a dynamical DE signal}. Nevertheless, both DESI BAO and FS modelling prefer a lower value of the (flat) $\Lambda$CDM parameter $\Omega_m$ relative to the Planck \cite{Planck:2018vyg}. In combination with external datasets, this still suffices to produce a statistically significant dynamical DE signal. 

DESI has recently released DR2 BAO results \cite{DESI:2025zgx}, showing that the combination DR2 BAO+CMB prefers dynamical DE at $3.1 \sigma$. As remarked in \cite{DESI:2024mwx, Colgain:2024xqj, Dinda:2024kjf, Wang:2024pui, Chudaykin:2024gol, Liu:2024gfy, Vilardi:2024cwq, Sapone:2024ltl}, there are noticeable fluctuations in both DR1 and DR2 BAO, especially in luminious red galaxies (LRG), which primarily drive the dynamical DE signal in DESI data alone. When combined with CMB and Type I supernovae (SNe) datasets, the overall discrepancy in $\Omega_m$ between the datasets increases the significance \cite{DESI:2025zgx}.  

Given that there is \textit{explicitly} a dynamical DE signal in BAO \cite{DESI:2024mwx, DESI:2024hhd, DESI:2025zgx} ($w_0 > -1$), but \textit{implicitly} no dynamical DE signal in FS modelling \cite{DESI:2024jis}, there is a degree of confusion. What compounds the confusion is the omission of explicit FS modelling + BAO entries for the Chevallier-Polarski-Linder (CPL) $w_0 w_a$CDM model \cite{Chevallier:2000qy, Linder:2002et} in Table 2 of the DESI DR1 FS modelling paper \cite{DESI:2024hhd}: Only FS + BAO in combination with both CMB and SNe datasets are considered. FS + BAO in combination with CMB alone is not. The reason for the omission is documented projected effects in Bayesian posteriors in the absence of SNe data \cite{DESI:2024hhd} (see \cite{DESI:2025hao} for frequentist analysis).  

On the other hand, Table 3 of the DR1 BAO paper \cite{DESI:2024mwx} shows the analogous constraints for BAO. Thus, the point of our letter is to make it explicit that fits of the CPL model to DESI FS modelling and BAO constraints \cite{DESI:2024jis}, both with an without CMB, lead to results consistent with $\Lambda$CDM within $1 \sigma$. This addresses a gap in the current literature. The upshot of this outcome is that one can confirm late-time accelerated expansion today, $q_0 < 0$, in DESI FS modelling + BAO alone confronted to the CPL model at $\gtrsim 3.4 \sigma$. As highlighted initially in \cite{Colgain:2025nzf}, the same cannot currently be said for DESI DR1 and DR2 BAO. 

\section{Analysis}
We employ a technique we have been using in previous papers \cite{Colgain:2024xqj, Colgain:2024mtg}, allowing one to map the CPL model back into the $\Lambda$CDM model at a given redshift $z_i$:
\begin{equation}
\label{eq:ratio}
        \frac{D_{M}(z_i)}{D_H(z_i)} = E(z_i) \int_0^{z_i} \frac{\dd z}{E(z)}.
\end{equation}
In the above, the left hand side is computed for the CPL model,
\begin{equation}
\label{eq:CPL}
\begin{split}
    D_{M}(z):= c \int_{0}^{z} \frac{\dd z^{\prime}}{H(z^{\prime})} ,&\qquad D_{H}(z) := \frac{c}{H(z)}\\
\hspace*{-3mm}H^2(z)=H_0^2\bigg[\Omega_m (1+z)^3+ (1-&\Omega_m)(1+z)^{3(1+\omega_0+\omega_a)} e^{-\frac{3\omega_a z}{1+z}}\bigg],
\end{split}
\end{equation} 
while the right hand side of \eqref{eq:ratio} is computed for the $\Lambda$CDM model for which  
\begin{equation}
    E^2(z)= 1-\tilde{\Omega}_m +\tilde{\Omega}_m (1+z)^3.
\end{equation}
Note that the matter density of the CPL model $\Omega_m$ is distinct from that of the $\Lambda$CDM model $\tilde{\Omega}_m$. Henceforth, we denote the $\Lambda$CDM matter density parameter with a tilde to avoid confusion. Observe also that $H_0$ drops out from the left hand side, so we simply fix $H_0 = 70$ km/s/Mpc. As a result, one is fitting only the $(\Omega_m, w_0, w_a)$ parameters from the CPL model. {We employ \eqref{eq:ratio} only at low redshift where any contribution from radiation is negligible.} We will discuss how one incorporates CMB in due course.

The mapping in \eqref{eq:ratio} can be applied beyond CPL more generally to any FLRW model on the left hand side to map it into the $\Lambda$CDM parameter $\tilde{\Omega}_m$. If there are no deviations from $\Lambda$CDM behaviour, one recovers a constant $\tilde{\Omega}_m$ at all redshifts probed. There is overlap with the $Om(z)$ diagnostic \cite{Sahni:2008xx}, but $Om(z)$ is usually continuous, necessitating a reconstruction of $E(z):= H(z)/H_0$, whereas \eqref{eq:ratio} begins from the cosmological distances at discrete redshifts typically constrained more directly by observations, e. g. BAO. In addition, $Om(z)$ typically blows up at lower $z$ (one can try propagating redshift errors to ameliorate this), whereas  \eqref{eq:ratio} propagates errors in both the numerator and denominator and cannot blow up at lower $z$. One could simply equate $H(z)^{\textrm{FLRW}} = H(z)^{\Lambda\textrm{CDM}}$, but then one would need to assume that $H_0^{\textrm{FLRW}} = H^{\Lambda \textrm{CDM}}_0$ to get a constraint on $\tilde{\Omega}_m$. Here, we do not need to assume $H_0^{\textrm{FLRW}} = H^{\Lambda \textrm{CDM}}_0$, since $H_0$ drops out in the ratio.

\begin{table}[htb]
\centering 
\begin{tabular}{|c|c|c|}
\hline
\rule{0pt}{3ex} \textbf{Tracer} & \textbf{$z_{\textrm{eff}}$} & \textbf{$\tilde\Omega_m$} \\
\hline\hline 
\rule{0pt}{3ex} BGS & $0.295$ & $0.284 \pm 0.024$ \\
\hline
\rule{0pt}{3ex} LRG1 & $0.510$ & $0.307^{+0.018}_{-0.020}$ \\
\hline
\rule{0pt}{3ex} LRG2 & $0.706$ & $0.287 \pm 0.020$ \\
\hline
\rule{0pt}{3ex} LRG3 & $0.919$ & $0.304 \pm 0.023$ \\
\hline
\rule{0pt}{3ex} ELG2 & $1.317$ & $0.310^{+0.027}_{-0.034}$ \\
\hline
\rule{0pt}{3ex} QSO & $1.491$ & $0.314^{+0.029}_{-0.039}$ \\
\hline
\end{tabular}
\caption{Constraints on $\tilde\Omega_m$ from FS modelling + BAO from different tracers at different effective redshifts. Redshifts and constraints are reproduced from Table 1 and Table 10 of \cite{DESI:2024jis}.}
\label{tab:constraints}
\end{table}

In our analysis we use of the constraints on $\tilde{\Omega}_m$ provided by the DESI collaboration from FS modelling + BAO \cite{DESI:2024jis} in Table \ref{tab:constraints}. One could use the constraints from FS modelling alone, but it will not change the conclusions, since BAO does not have a strong bearing on FS modelling (see Table 10 of \cite{DESI:2024jis}). We define a log-likelihood $\mathcal{L}( \Omega_m, w_0, w_a)$ with input parameters $(\Omega_m, w_0, w_a)$ from the CPL model. For each $z_i \in z_{\textrm{eff}}$ in Table \ref{tab:constraints}, we solve equation (\ref{eq:ratio}) to identify the corresponding $\tilde{\Omega}_m (z_i)$ value from the $\Lambda$CDM model. From there, we define the log-likelihood 
\begin{equation}
\label{eq:likelihood}
    \log \mathcal{L} (\Omega_m, w_0, w_a) = - \frac{1}{2} \chi^2 = - \frac{1}{2} \sum_{i} \frac{(\tilde{\Omega}_m (z_i) - \tilde\Omega^i_m)^2}{\sigma_{\tilde{\Omega}^i_m}^2}, 
\end{equation}
where $\tilde\Omega^i_m$ denote the central value of the $\tilde\Omega_m$ values and $\sigma_{\tilde{\Omega}^i_m}$ denotes the errors in Table \ref{tab:constraints}. 
Given the antisymmetric errors in Table \ref{tab:constraints}, if $\tilde{\Omega}_{m}(z_i) > \tilde\Omega_m^i$ we use the upper error, and \textit{vice versa}. This ensures that our log-likelihood properly takes account of the differences in antisymmetric errors. Note, we are unaware of any work in the literature fitting model A to model B constraints by rewriting it as model B. This may be a novel workaround for model comparison.

\begin{figure}[htb]
   \centering
\includegraphics[width=80mm]{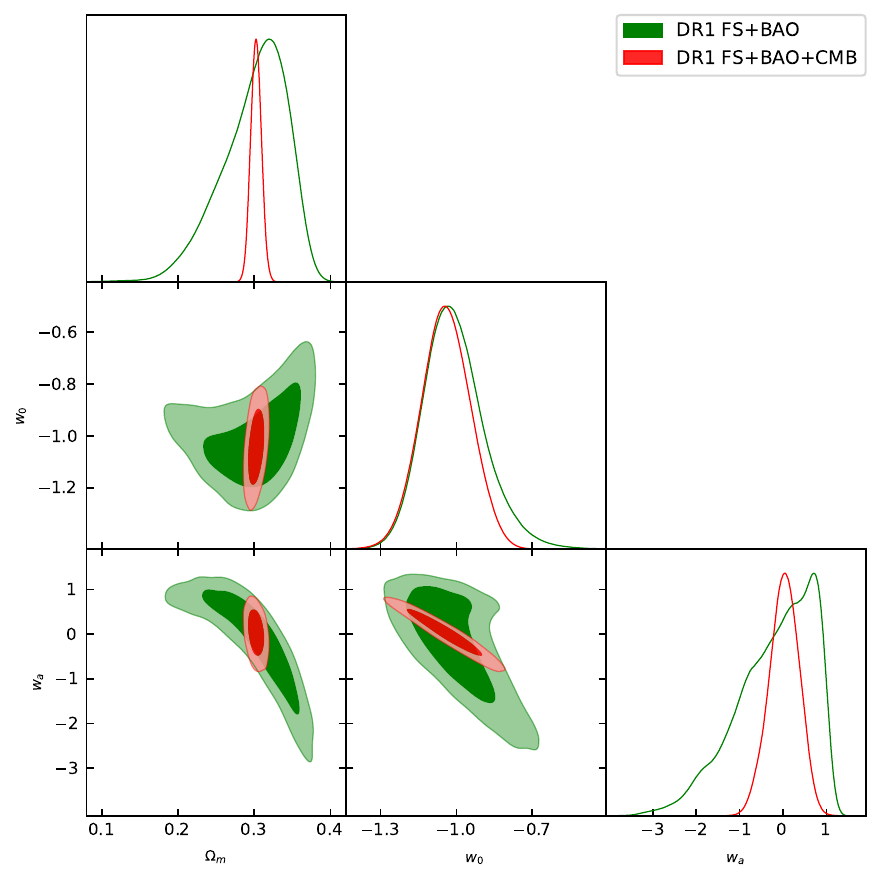}
\caption{CPL parameter posteriors from DESI DR1 FS modelling + BAO with and without CMB.}
\label{fig:corner} 
\end{figure}

Once the log-likelihood is defined, one marginalises over the CPL parameters $(\Omega_m, w_0, w_a)$ with Markov Chain Monte Carlo (MCMC) to identify the parameters that best fit the DESI FS + BAO constraints. We employ \textit{emcee} \cite{Foreman-Mackey:2012any} with the DESI priors,  $w_0 \in [-3, 1], w_a \in [-3, 2]$ and $w_0 + w_a < 0$ \cite{DESI:2024mwx}. The result of this exercise is presented in Fig. \ref{fig:corner} and Table \ref{tab:results}, where we have used \textit{getdist} \cite{Lewis:2019xzd} to plot the posteriors. The ($w_0, w_a$) values are within $1 \sigma$ of $\Lambda$CDM, $(w_0, w_a) = (-1, 0)$, so there is no trace of dynamical DE, as expected. There is a concern that our methodology, namely fitting the CPL model through the $\Lambda$CDM model to $\Lambda$CDM constraints, risks washing out a dynamical DE signal even if one is present. In the appendix, we perform a consistency check to show that we can recover the dynamical DE signal in DESI DR1 BAO using both Bayesian and frequentist methods. Our MCMC posteriors are impacted by priors and projection/volume effects, but in a frequentist analysis it is clear that $w_0 > -1$. 

\begin{table*}
\centering 
\begin{tabular}{|c|c|c|c|c|}
\hline
\rule{0pt}{3ex} \textbf{Data} & \textbf{$\Omega_m$} & \textbf{$w_0$} & \textbf{$w_a$} & $q_0$\\
\hline\hline 
\rule{0pt}{3ex} FS + BAO & $0.307^{+0.036}_{-0.051}$ & $-1.02^{+0.12}_{-0.11}$ & $-0.03^{+0.76}_{-1.1}$ & $-0.58^{+0.17}_{-0.13}$ \\
\hline
\rule{0pt}{3ex} FS + BAO + CMB & $0.3025^{+0.0070}_{-0.0069}$ & $-1.042^{+0.099}_{-0.097}$ & $0.03^{+0.34}_{-0.36}$ & $-0.59 \pm 0.11$ \\
\hline
\end{tabular}
\caption{Constraints on the $\Lambda$CDM parameter $\Omega_m$ from BAO+FS Modelling from different tracers at different effective redshifts. Redshifts and constraints reproduced from Table 1 and Table 10 of \cite{DESI:2024jis}.}
\label{tab:results}
\end{table*}

\begin{figure}[htb]
   \centering
\includegraphics[width=80mm]{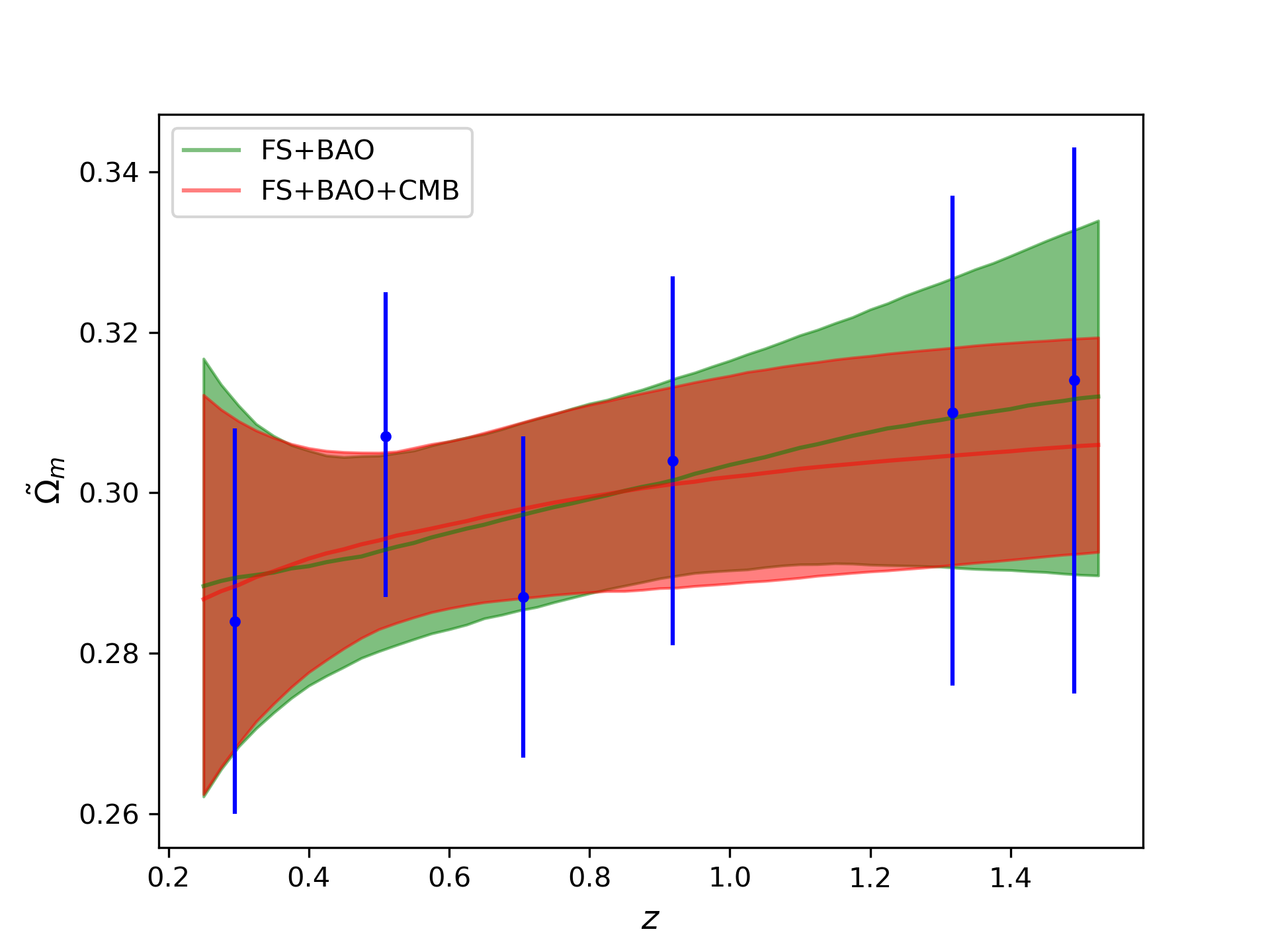}
\caption{Reconstructed $\tilde{\Omega}_m$ from DESI DR1 FS modelling + BAO with and without CMB.}
\label{fig:OM} 
\end{figure}

At this stage, there is a further consistency check one can perform. It is evident from the $\tilde\Omega_m$ values in Table \ref{tab:constraints} and the blue constraints in Fig. \ref{fig:OM} that there is a mild increasing trend in the $\tilde\Omega_m$ central values. We expect to see this in the best fit CPL model when it is mapped back to $\Lambda$CDM. Given the MCMC chain, one can use \eqref{eq:ratio} and redshifts in the range $ z \in [0.25, 1.5]$ separated by a uniform $\Delta z = 0.025$ to reconstruct a distribution of $\tilde{\Omega}_m$ values at each redshift. One then isolates $16^{\textrm{th}}$ and $84^{\textrm{th}}$ percentiles of $\tilde{\Omega}_m$ at each redshifts as the limits of the $68 \%$ confidence intervals and the median as the central value. What one expects to find through the consistency check is that the increasing trend evident in the blue constraints is mirrored in the resulting confidence interval band. As can be seen from Fig. \ref{fig:OM}, one sees this feature in the green band.   

The next step is to incorporate CMB to see if this makes a difference to conclusions. Here, we follow the DESI collaboration and introduce the Gaussian priors on $(\theta_*, \omega_b, \omega_m)$ from appendix A of \cite{DESI:2025zgx}, where we define $\theta_* = r_*/D_M(z_*)$, $\omega_b = \Omega_b h^2$ and $\omega_m = \Omega_m h^2$. $\Omega_b$ is the baryonic matter density parameter, $r_*$ denotes the comoving sound horizon at last scattering and $h = H_0/(100 \textrm{ km/s/Mpc})$. This reintroduces $H_0$, which drops out of the log-likelihood in (\ref{eq:likelihood}), but is relevant for the CMB constraints. Furthermore, $\Omega_b$ appears as a second additional parameter. Finally, we fix $z_* = 1090$ and introduce a radiation sector in the CPL model (\ref{eq:CPL}) with the standard $a^{-4}$ scaling with fixed coefficient $\Omega_r = 4.18 \times 10^{-5}/h^2$, where $h:= H_0/[100 \textrm{km/s/Mpc}]$. We add the CMB log-likelihood dependent on $v = (\theta_*, \omega_b, \omega_m)$ to the log-likelihood for FS modelling + BAO, which results in a log-likelihood that depends on $(H_0, \Omega_m, \Omega_b, w_0, w_a)$: 
\begin{equation}
\begin{split}
     \log \mathcal{L} (H_0, \Omega_m, \Omega_b, w_0, w_a) = & - \frac{1}{2} \sum_{i} \frac{(\tilde{\Omega}_m (z_i) - \tilde\Omega^i_m)^2}{\sigma_{\tilde{\Omega}^i_m}^2}  \\ &- \frac{1}{2} \Delta v \cdot C^{-1} \cdot \Delta v,
\end{split}
\end{equation}
where we have defined $\Delta v = v - v_{\textrm{theory}}$. Expressions for $v$ and $C$ can be found in appendix A of \cite{DESI:2025zgx} and one calculates $v_{\textrm{theory}} = (\theta_*, \omega_b, \omega_m)$ from the log-likelihood input parameters.
Marginalising over the parameters through MCMC, while isolating $(\Omega_m, w_0, w_a)$ for comparison, one gets the result in Fig. \ref{fig:corner}, where the corresponding $68 \%$ confidence intervals can be found in Table \ref{tab:results}. The consistency check of reconstructing $\tilde{\Omega}_m$ appears in Fig. \ref{fig:OM}, where we see that the red (brownish) confidence interval band shows the expected increasing redshift trend.

A number of comments are in order. First, comparing FS + BAO with and without CMB from Table \ref{tab:results} one can see the difference CMB makes. CMB greatly increases the $\Omega_m$ precision, decreases the $w_a$ errors by a factor of $2-3$, but there is no great increase in $w_0$ precision. This is also reflected in the reconstructed $\tilde{\Omega}_m$ in Fig. \ref{fig:OM}. What one sees is that the green and red (brownish) confidence intervals show little difference at lower redshifts, where the DE sector parameterised by $(w_0, w_a)$ is most relevant, whereas at higher redshifts in the matter dominated regime, the confidence intervals contract appreciably. One also notes that the CMB is forcing the reconstruction to a relatively lower canonical $\tilde{\Omega}_m \sim 0.3$ value at higher redshifts. Secondly, whether one employs CMB constraints or not, we find $(w_0, w_a)$ values that are within $1 \sigma$ of $\Lambda$CDM.  

Finally, we come to a key remark. In \cite{Colgain:2025nzf} it was noted that DESI DR1 \cite{DESI:2024mwx} and DR2 BAO \cite{DESI:2025zgx} when confronted to the CPL model fails to confirm late-time accelerated expansion today, which is characterised by a negative deceleration parameter:\footnote{See also \cite{Wang:2024rjd} for an earlier implicit observation that $w_0 > -\frac{1}{3}$ and \cite{Wang:2025bkk, Wang:2025vtw} for later explicit observations that $q_0 < 0$. Note that as \eqref{q0-CPL} shows, $w_0  < -\frac{1}{3}$ is a necessary condition that implies $q_0 < 0$ only for $\Omega_m = 0$.}
\begin{equation}\label{q0-CPL}
q_0 = \frac{1}{2} \left[ 1+3 w_0 (1-\Omega_m) \right] < 0.     
\end{equation}

\begin{figure}[htb]
   \centering
\includegraphics[width=70mm]{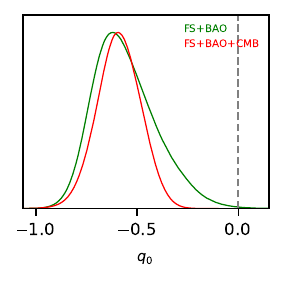}
\caption{Reconstructed deceleration parameter $q_0$ posteriors from fits of the CPL model to DESI DR1 FS modelling + BAO with and without CMB. }
\label{fig:q0} 
\end{figure}

In contrast, from Table \ref{tab:results} and Fig. \ref{fig:q0}, we see that FS + BAO confirms late-time accelerated expansion at $ 3.4 \sigma$ without CMB and at $5.4 \sigma$ with CMB. This highlights a key difference in the physical implications of BAO alone compared to FS + BAO. 

\section{Outlook}
Where will future DESI results take us? What should be clear is that DESI BAO and DESI FS modelling results need to converge. It is evident from analysis in \cite{Colgain:2025nzf} that DESI BAO shows good consistency between DR1 and DR2 at higher redshifts, but that fluctuations are still present in LRG and potentially the lowest redshift emission line galaxy (ELG) bin. More concretely, while LRG1 was primarily responsible for the dynamical DE signal in DR1 BAO data alone \cite{Colgain:2024xqj}, in DR2 BAO, LRG2 is the main driver of the dynamical DE signal \cite{Colgain:2025nzf}. When data points move about to this extent, the risk is that fluctuations are present. 

The point of this letter is to make the implicit explicit and provide results that were omitted in Table 2 of \cite{DESI:2024hhd}, namely constraints on the CPL model from DESI DR1 FS + BAO and FS + BAO + CMB. Projection effects were used to justify the omission of the results on the grounds that the mean and mode of posteriors did not agree \cite{DESI:2024hhd}. From a mathematical perspective, this is puzzling, since if the $\Lambda$CDM model is well constrained in redshift bins, this has implications for the CPL dark energy model;  (\ref{eq:ratio}) is a mapping between any late-universe FLRW cosmology and $\Lambda$CDM. With additional nuisance parameters beyond cosmological parameters, projection effects can easily arise due to degeneracies between the parameters.

We have demonstrated that there is no hint of a dynamical DE signal in FS modelling, even when combined with CMB. To this end, we have employed a mapping between the CPL model and $\Lambda$CDM. This strategy of mapping model A to model B to fit model A to model B constraints may be a novel workaround; we are unaware of any examples. We have tested our method in the appendix. However, given one can run a horizontal constant $\tilde{\Omega}_m$ line through the blue constraints in Fig. \ref{fig:OM}, this is the expected and obvious result. Finally, while DESI DR1 and DR2 BAO fail to confirm late-time accelerated expansion today \cite{Colgain:2025nzf}, we see that any combination with FS modelling does this in excess of $3 \sigma$. 

The interesting question now is, assuming BAO converges to FS modelling, what will happen when the blue constraints in Fig. \ref{fig:OM} shrink as FS modelling + BAO results are upgraded from DR1 to DR2? If the central values do not shift, this will leave an increasing $\tilde{\Omega}_m$ trend with redshift. We remind the reader that in \cite{Colgain:2024xqj} it was conjectured that an increasing $\tilde{\Omega}_m$ signal would emerge from DESI data, in particular BAO. The flip side of an increasing matter density parameter with redshift in the $\Lambda$CDM model is a decreasing Hubble constant $H_0$ with redshift in the $\Lambda$CDM model \footnote{See \cite{Colgain:2024ksa} for a manifestation of the anti-correlated trends.}, thereby corroborating Hubble tension, a discrepancy in $H_0$ between the early (high redshift) and late (low redshift) Universe \cite{DiValentino:2025sru}. More generally, see \cite{Akarsu:2024qiq} for a review of earlier observations of redshift-dependent $\Lambda$CDM parameters - a hallmark of model breakdown - in different observables. See also \cite{Mukherjee:2024pcg} for recent relevant observation that redshift-dependent $\Lambda$CDM fitting parameters\footnote{Concretely, $H_0$ decreasing with redshift, $\Omega_m$ increasing with redshift and $\sigma_8/S_8$ increasing with redshift.} can be found in SDSS data. 

Finally,  great care is required when combining BAO and SNe datasets. One needs to check that BAO and SNe agree on cosmological distances in overlapping redshift ranges, e. g. \cite{Colgain:2024mtg}. Two groups \cite{Teixeira:2025czm, Afroz:2025iwo} have recently reported a breakdown in the distance duality relation when DESI BAO is combined with SNe. The physical implications of such a breakdown are so profound (giving up conservation of photon propogation in a metric theory of gravity) that systematics must be present.   

\section*{Acknowledgements}
This article/publication is based upon work from COST Action CA21136 – “Addressing observational tensions in cosmology with systematics and fundamental physics (CosmoVerse)”, supported by COST (European Cooperation in Science and Technology). 

\appendix 

\section{Methodology Check}

In the main text we employed a log-likelihood that converts the CPL model into a $\Lambda$CDM model before fitting the resulting $\Lambda$CDM model to DESI DR1 FS + BAO $\Lambda$CDM constraints. For FS + BAO both with and without CMB we found $w_0 =-1$ within $1 \sigma$. One concern one may have is that any dynamical DE signal may be washed out by our methodology. Thus, the check that needs to be performed is to make sure that we can recover a $w_0 > -1$ (alternatively $w_a < 0$) signal from a dataset with a \textit{bona fide} dynamical DE signal. For this test we will make use of DESI DR1 BAO, which confronted to the CPL model directly returns $w_0 > -1$ at in excess of $1 \sigma$.

\begin{table}[htb]
\centering 
\begin{tabular}{|c|c|}
\hline
\rule{0pt}{3ex} \textbf{$z_{\textrm{eff}}$} & \textbf{$\tilde\Omega_m$} \\
\hline\hline 
\rule{0pt}{3ex} $0.51$ & $0.67^{+0.18}_{-0.17}$ \\
\hline
\rule{0pt}{3ex} $0.71$ & $0.219^{+0.087}_{-0.069}$ \\
\hline
\rule{0pt}{3ex} $0.93$ & $0.276^{+0.053}_{-0.047}$ \\
\hline
\rule{0pt}{3ex} $1.32$ & $0.345^{+0.11}_{-0.078}$ \\
\hline
\rule{0pt}{3ex} $2.33$ & $0.375^{+0.088}_{-0.069}$ \\
\hline
\end{tabular}
\caption{Constraints on $\tilde\Omega_m$ from Table I of  \cite{Colgain:2024xqj} based on DESI DR1 BAO.}
\label{tab:check_constraints}
\end{table}

\begin{figure}[htb]
   \centering
\includegraphics[width=80mm]{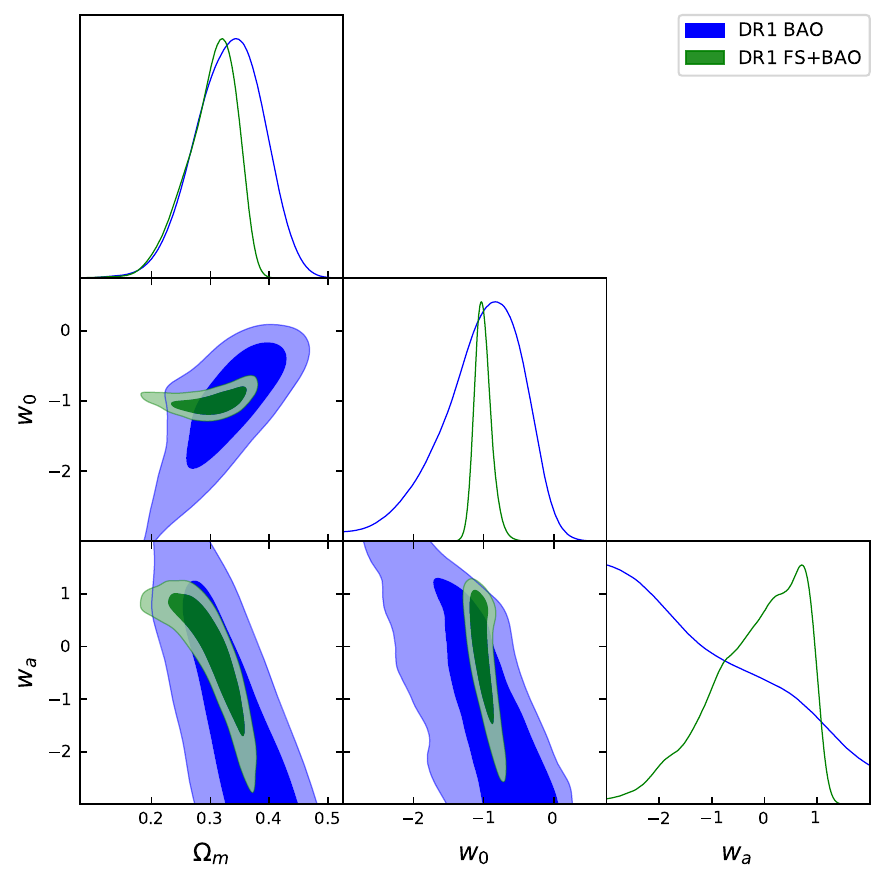}
\caption{CPL parameter posteriors for DESI DR1 FS + BAO and DR1 BAO confronted to $\Lambda$CDM constraints on $\tilde{\Omega}_m$.}
\label{fig:FS_BAO_corner} 
\end{figure}

To begin, we need constraints on the $\Lambda$CDM parameter $\tilde{\Omega}_m (z_i)$ at effective redshift $z_i$. Here we can import Table I of \cite{Colgain:2024xqj} where the DESI DR1 BAO constraints on $D_M(z_i)$ and $D_H(z_i)$ were converted into direct constraints on $\tilde{\Omega}_m(z_i)$. We reproduce the constraints in Table \ref{tab:check_constraints}, where it should be evident that one can cannot interpolate a constant $\tilde{\Omega}_m$ through the error bars (see Fig. 1 of \cite{Colgain:2024xqj}). As a result, there is a hint of a $\Lambda$CDM deviation that is interpretable as dynamical DE. We now replace the constraints in Table \ref{tab:constraints} with the constraints in Table \ref{tab:check_constraints} to see what difference it makes. In Fig. \ref{fig:FS_BAO_corner} we present a corner plot where the green posterior is the same as Fig. \ref{fig:corner}.

Comparing the green (DR1 FS + BAO) and blue posteriors (DR1 BAO), we see key differences. First, with the DESI priors, the green posterior is constrained in the $(w_0, w_a)$-plane, whereas the blue posterior is not. This is not surprising as the fractional errors in Table \ref{tab:constraints} are smaller than Table \ref{tab:check_constraints}; FS modelling + BAO constrains the $\Lambda$CDM model much better than BAO alone. We note that $w_0 = -0.97^{+0.50}_{-0.66}$ at $68 \%$ credible level ($1 \sigma$) is consistent with $w_0 =-1$ within $1 \sigma$, but the maximum of the log-likelihood from the MCMC chain occurs at $w_0 = -0.31$, which is outside the credible interval. This points to a projection effect. Moreover, given the anti-correlation between $w_0$ and $w_a$ in the blue posterior, it should be clear that relaxing the lower bound $w_a \geq -3$ will drag the $w_0$ posterior to arbitrarily larger values that depend on the $w_a$ prior. In summary, there is evidence for dynamical DE in the $w_a$ posterior, but there is no signal in the $w_0$ posterior. This is due to the priors and marginalsation, as we will now demonstrate.

\begin{figure}[htb]
   \centering
\includegraphics[width=80mm]{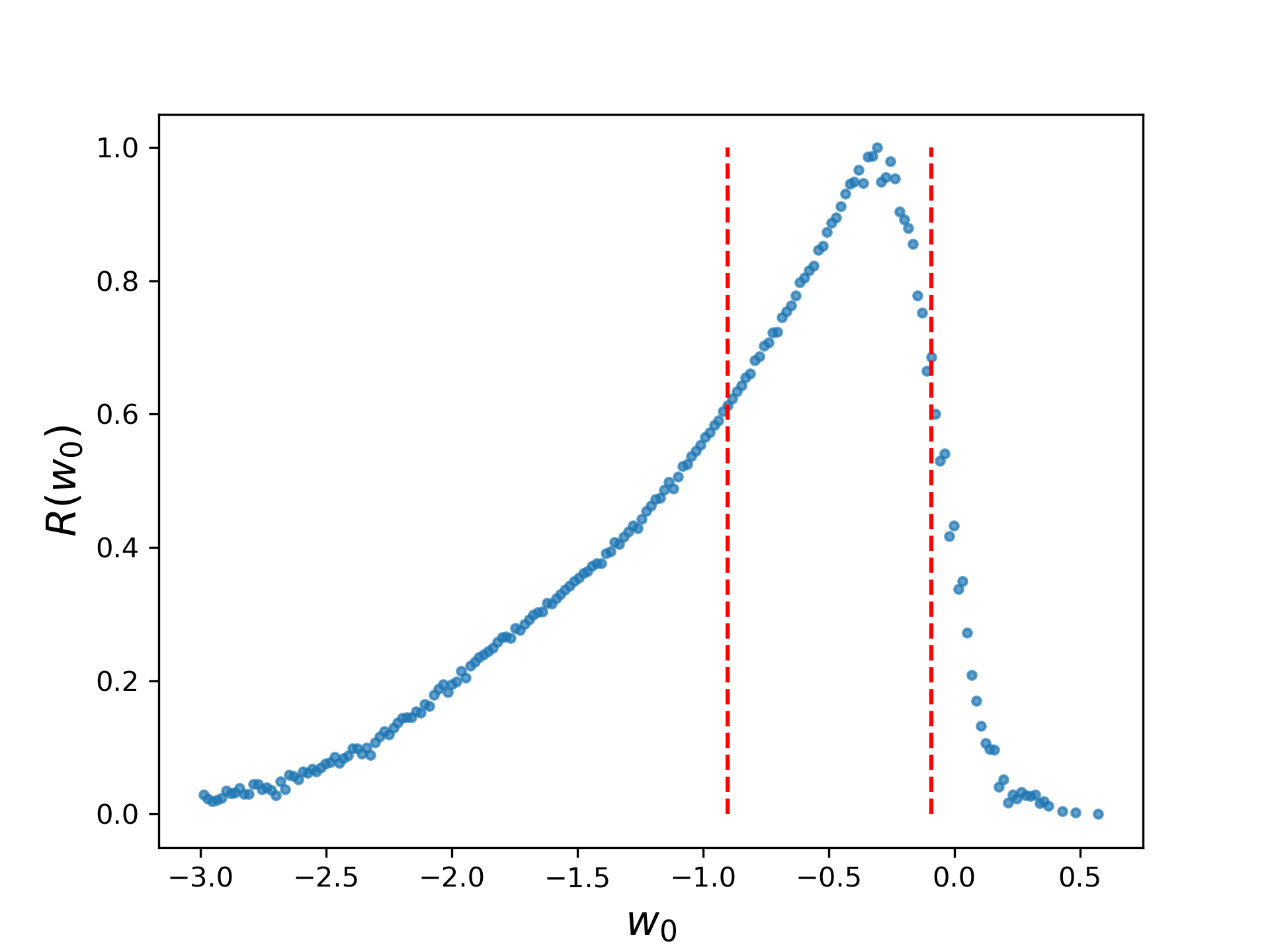}
\caption{$w_0$ profile likelihood for DESI DR1 BAO with log-likelihood in \eqref{eq:likelihood}. Red dashed lines denote $68 \%$ confidence intervals estimated through Wilks' theorem.}
\label{fig:profile} 
\end{figure}

One can get a second perspective on this through frequentist profile likelihood methods following \cite{Gomez-Valent:2022hkb, Colgain:2023bge}, where one bins the MCMC chain in $w_0$ bins. Frequentist methods are less prone to the impact of priors and projection effects. We refer the reader to the above references for the details, but the main idea is to define the profile likelihood ratio:
\begin{equation}
\label{eq:R}
    R(w_0) = \exp \left( -\frac{1}{2} ( \chi^2_{\textrm{min}} (w_0) - \chi^2_{\textrm{min}})\right),  
\end{equation}
where $\chi^2_{\textrm{min}} (w_0)$ is the minimum value of the $\chi^2$ for the MCMC configurations in the bin centred on $w_0$ and $\chi^2_{\textrm{min}}$ is the global minimum for all MCMC configurations. We present $R(w_0)$ in Fig. \ref{fig:profile}. Given $R(w_0)$ one can get a $68 \%$  confidence interval from Wilks' theorem \cite{Wilks:1938dza} through identifying the range of $w_0$ values with $\Delta \chi^2 \leq 1 \Leftrightarrow R(w_0) \geq e^{-\frac{1}{2}} \approx 0.607$. Strictly speaking, the theorem only holds for profile likelihoods close to Gaussian, but it is a quick way to get an indicative result. The resulting constraint on $w_0$ is $w_0 = -0.31^{+0.21}_{-0.60}$, thereby confirming that $w_0 > -1$ beyond $1 \sigma$. There is noise evident in $R(w_0)$ in Fig. \ref{fig:profile}, but this comes about because MCMC is a poor optimiser and the $R(w_0)$ dots converge to a smooth $R(w_0)$ curve from below.\footnote{See Fig. 4 of \cite{Colgain:2024clf} for a comparison between profile likelihoods based on gradient decent and binning the MCMC chain. As explained by Trotta \cite{Trotta:2017wnx}, both methods are acceptable.}

What Fig. \ref{fig:profile} demonstrates is that marginalisation over the DR1 BAO posterior in Fig. \ref{fig:FS_BAO_corner} drags the projected 1D $w_0$ posterior back closer to $w_0 = -1$ through a volume/projection effect. Despite this difficulty in seeing the dynamical DE signal in Fig. \ref{fig:FS_BAO_corner}, it is clear that $w_a < -3$ values are preferred and this dynamical DE signal is not evident in the green posterior. This is a like-for-like comparison with the same method.

\bibliography{refs}

\begin{thebibliography}{34}%
\makeatletter
\providecommand \@ifxundefined [1]{%
 \@ifx{#1\undefined}
}%
\providecommand \@ifnum [1]{%
 \ifnum #1\expandafter \@firstoftwo
 \else \expandafter \@secondoftwo
 \fi
}%
\providecommand \@ifx [1]{%
 \ifx #1\expandafter \@firstoftwo
 \else \expandafter \@secondoftwo
 \fi
}%
\providecommand \natexlab [1]{#1}%
\providecommand \enquote  [1]{``#1''}%
\providecommand \bibnamefont  [1]{#1}%
\providecommand \bibfnamefont [1]{#1}%
\providecommand \citenamefont [1]{#1}%
\providecommand \href@noop [0]{\@secondoftwo}%
\providecommand \href [0]{\begingroup \@sanitize@url \@href}%
\providecommand \@href[1]{\@@startlink{#1}\@@href}%
\providecommand \@@href[1]{\endgroup#1\@@endlink}%
\providecommand \@sanitize@url [0]{\catcode `\\12\catcode `\$12\catcode `\&12\catcode `\#12\catcode `\^12\catcode `\_12\catcode `\%12\relax}%
\providecommand \@@startlink[1]{}%
\providecommand \@@endlink[0]{}%
\providecommand \url  [0]{\begingroup\@sanitize@url \@url }%
\providecommand \@url [1]{\endgroup\@href {#1}{\urlprefix }}%
\providecommand \urlprefix  [0]{URL }%
\providecommand \Eprint [0]{\href }%
\providecommand \doibase [0]{http://dx.doi.org/}%
\providecommand \selectlanguage [0]{\@gobble}%
\providecommand \bibinfo  [0]{\@secondoftwo}%
\providecommand \bibfield  [0]{\@secondoftwo}%
\providecommand \translation [1]{[#1]}%
\providecommand \BibitemOpen [0]{}%
\providecommand \bibitemStop [0]{}%
\providecommand \bibitemNoStop [0]{.\EOS\space}%
\providecommand \EOS [0]{\spacefactor3000\relax}%
\providecommand \BibitemShut  [1]{\csname bibitem#1\endcsname}%
\let\auto@bib@innerbib\@empty
\bibitem [{\citenamefont {Adame}\ \emph {et~al.}(2025)\citenamefont {Adame} \emph {et~al.}}]{DESI:2024mwx}%
  \BibitemOpen
  \bibfield  {author} {\bibinfo {author} {\bibfnamefont {A.~G.}\ \bibnamefont {Adame}} \emph {et~al.} (\bibinfo {collaboration} {DESI}),\ }\href {\doibase 10.1088/1475-7516/2025/02/021} {\bibfield  {journal} {\bibinfo  {journal} {JCAP}\ }\textbf {\bibinfo {volume} {02}},\ \bibinfo {pages} {021} (\bibinfo {year} {2025})},\ \Eprint {http://arxiv.org/abs/2404.03002} {arXiv:2404.03002 [astro-ph.CO]} \BibitemShut {NoStop}%
\bibitem [{\citenamefont {{Adame}}\ \emph {et~al.}(2024{\natexlab{a}})\citenamefont {{Adame}} \emph {et~al.}}]{DESI:2024hhd}%
  \BibitemOpen
  \bibfield  {author} {\bibinfo {author} {\bibfnamefont {A.~G.}\ \bibnamefont {{Adame}}} \emph {et~al.} (\bibinfo {collaboration} {{DESI}}),\ }\href {\doibase 10.48550/arXiv.2411.12022} {\bibfield  {journal} {\bibinfo  {journal} {arXiv e-prints}\ ,\ \bibinfo {eid} {arXiv:2411.12022}} (\bibinfo {year} {2024}{\natexlab{a}})},\ \Eprint {http://arxiv.org/abs/2411.12022} {arXiv:2411.12022 [astro-ph.CO]} \BibitemShut {NoStop}%
\bibitem [{\citenamefont {Abdul~Karim}\ \emph {et~al.}(2025)\citenamefont {Abdul~Karim} \emph {et~al.}}]{DESI:2025zgx}%
  \BibitemOpen
  \bibfield  {author} {\bibinfo {author} {\bibfnamefont {M.}~\bibnamefont {Abdul~Karim}} \emph {et~al.} (\bibinfo {collaboration} {DESI}),\ }\href@noop {} {\  (\bibinfo {year} {2025})},\ \Eprint {http://arxiv.org/abs/2503.14738} {arXiv:2503.14738 [astro-ph.CO]} \BibitemShut {NoStop}%
\bibitem [{\citenamefont {{Adame}}\ \emph {et~al.}(2024{\natexlab{b}})\citenamefont {{Adame}} \emph {et~al.}}]{DESI:2024jis}%
  \BibitemOpen
  \bibfield  {author} {\bibinfo {author} {\bibfnamefont {A.~G.}\ \bibnamefont {{Adame}}} \emph {et~al.} (\bibinfo {collaboration} {DESI}),\ }\href {\doibase 10.48550/arXiv.2411.12021} {\bibfield  {journal} {\bibinfo  {journal} {arXiv e-prints}\ ,\ \bibinfo {eid} {arXiv:2411.12021}} (\bibinfo {year} {2024}{\natexlab{b}})},\ \Eprint {http://arxiv.org/abs/2411.12021} {arXiv:2411.12021 [astro-ph.CO]} \BibitemShut {NoStop}%
\bibitem [{\citenamefont {Aghanim}\ \emph {et~al.}(2020)\citenamefont {Aghanim} \emph {et~al.}}]{Planck:2018vyg}%
  \BibitemOpen
  \bibfield  {author} {\bibinfo {author} {\bibfnamefont {N.}~\bibnamefont {Aghanim}} \emph {et~al.} (\bibinfo {collaboration} {Planck}),\ }\href {\doibase 10.1051/0004-6361/201833910} {\bibfield  {journal} {\bibinfo  {journal} {Astron. Astrophys.}\ }\textbf {\bibinfo {volume} {641}},\ \bibinfo {pages} {A6} (\bibinfo {year} {2020})},\ \bibinfo {note} {[Erratum: Astron.Astrophys. 652, C4 (2021)]},\ \Eprint {http://arxiv.org/abs/1807.06209} {arXiv:1807.06209 [astro-ph.CO]} \BibitemShut {NoStop}%
\bibitem [{\citenamefont {Colg{\'a}in}\ \emph {et~al.}(2026)\citenamefont {Colg{\'a}in}, \citenamefont {Dainotti}, \citenamefont {Capozziello}, \citenamefont {Pourojaghi}, \citenamefont {Sheikh-Jabbari},\ and\ \citenamefont {Stojkovic}}]{Colgain:2024xqj}%
  \BibitemOpen
  \bibfield  {author} {\bibinfo {author} {\bibfnamefont {E.~{\'O}.}\ \bibnamefont {Colg{\'a}in}}, \bibinfo {author} {\bibfnamefont {M.~G.}\ \bibnamefont {Dainotti}}, \bibinfo {author} {\bibfnamefont {S.}~\bibnamefont {Capozziello}}, \bibinfo {author} {\bibfnamefont {S.}~\bibnamefont {Pourojaghi}}, \bibinfo {author} {\bibfnamefont {M.~M.}\ \bibnamefont {Sheikh-Jabbari}}, \ and\ \bibinfo {author} {\bibfnamefont {D.}~\bibnamefont {Stojkovic}},\ }\href {\doibase 10.1016/j.jheap.2025.100428} {\bibfield  {journal} {\bibinfo  {journal} {JHEAp}\ }\textbf {\bibinfo {volume} {49}},\ \bibinfo {pages} {100428} (\bibinfo {year} {2026})},\ \Eprint {http://arxiv.org/abs/2404.08633} {arXiv:2404.08633 [astro-ph.CO]} \BibitemShut {NoStop}%
\bibitem [{\citenamefont {Dinda}(2024)}]{Dinda:2024kjf}%
  \BibitemOpen
  \bibfield  {author} {\bibinfo {author} {\bibfnamefont {B.~R.}\ \bibnamefont {Dinda}},\ }\href {\doibase 10.1088/1475-7516/2024/09/062} {\bibfield  {journal} {\bibinfo  {journal} {JCAP}\ }\textbf {\bibinfo {volume} {09}},\ \bibinfo {pages} {062} (\bibinfo {year} {2024})},\ \Eprint {http://arxiv.org/abs/2405.06618} {arXiv:2405.06618 [astro-ph.CO]} \BibitemShut {NoStop}%
\bibitem [{\citenamefont {Wang}\ \emph {et~al.}(2024)\citenamefont {Wang}, \citenamefont {Lin}, \citenamefont {Ding},\ and\ \citenamefont {Hu}}]{Wang:2024pui}%
  \BibitemOpen
  \bibfield  {author} {\bibinfo {author} {\bibfnamefont {Z.}~\bibnamefont {Wang}}, \bibinfo {author} {\bibfnamefont {S.}~\bibnamefont {Lin}}, \bibinfo {author} {\bibfnamefont {Z.}~\bibnamefont {Ding}}, \ and\ \bibinfo {author} {\bibfnamefont {B.}~\bibnamefont {Hu}},\ }\href {\doibase 10.1093/mnras/stae2309} {\bibfield  {journal} {\bibinfo  {journal} {Mon. Not. Roy. Astron. Soc.}\ }\textbf {\bibinfo {volume} {534}},\ \bibinfo {pages} {3869} (\bibinfo {year} {2024})},\ \Eprint {http://arxiv.org/abs/2405.02168} {arXiv:2405.02168 [astro-ph.CO]} \BibitemShut {NoStop}%
\bibitem [{\citenamefont {Chudaykin}\ and\ \citenamefont {Kunz}(2024)}]{Chudaykin:2024gol}%
  \BibitemOpen
  \bibfield  {author} {\bibinfo {author} {\bibfnamefont {A.}~\bibnamefont {Chudaykin}}\ and\ \bibinfo {author} {\bibfnamefont {M.}~\bibnamefont {Kunz}},\ }\href {\doibase 10.1103/PhysRevD.110.123524} {\bibfield  {journal} {\bibinfo  {journal} {Phys. Rev. D}\ }\textbf {\bibinfo {volume} {110}},\ \bibinfo {pages} {123524} (\bibinfo {year} {2024})},\ \Eprint {http://arxiv.org/abs/2407.02558} {arXiv:2407.02558 [astro-ph.CO]} \BibitemShut {NoStop}%
\bibitem [{\citenamefont {{Liu}}\ \emph {et~al.}(2024)\citenamefont {{Liu}}, \citenamefont {{Wang}},\ and\ \citenamefont {{Zhao}}}]{Liu:2024gfy}%
  \BibitemOpen
  \bibfield  {author} {\bibinfo {author} {\bibfnamefont {G.}~\bibnamefont {{Liu}}}, \bibinfo {author} {\bibfnamefont {Y.}~\bibnamefont {{Wang}}}, \ and\ \bibinfo {author} {\bibfnamefont {W.}~\bibnamefont {{Zhao}}},\ }\href {\doibase 10.48550/arXiv.2407.04385} {\bibfield  {journal} {\bibinfo  {journal} {arXiv e-prints}\ ,\ \bibinfo {eid} {arXiv:2407.04385}} (\bibinfo {year} {2024})},\ \Eprint {http://arxiv.org/abs/2407.04385} {arXiv:2407.04385 [astro-ph.CO]} \BibitemShut {NoStop}%
\bibitem [{\citenamefont {{Vilardi}}\ \emph {et~al.}(2024)\citenamefont {{Vilardi}}, \citenamefont {{Capozziello}},\ and\ \citenamefont {{Brescia}}}]{Vilardi:2024cwq}%
  \BibitemOpen
  \bibfield  {author} {\bibinfo {author} {\bibfnamefont {S.}~\bibnamefont {{Vilardi}}}, \bibinfo {author} {\bibfnamefont {S.}~\bibnamefont {{Capozziello}}}, \ and\ \bibinfo {author} {\bibfnamefont {M.}~\bibnamefont {{Brescia}}},\ }\href {\doibase 10.48550/arXiv.2408.01563} {\bibfield  {journal} {\bibinfo  {journal} {arXiv e-prints}\ ,\ \bibinfo {eid} {arXiv:2408.01563}} (\bibinfo {year} {2024})},\ \Eprint {http://arxiv.org/abs/2408.01563} {arXiv:2408.01563 [astro-ph.CO]} \BibitemShut {NoStop}%
\bibitem [{\citenamefont {{Sapone}}\ and\ \citenamefont {{Nesseris}}(2024)}]{Sapone:2024ltl}%
  \BibitemOpen
  \bibfield  {author} {\bibinfo {author} {\bibfnamefont {D.}~\bibnamefont {{Sapone}}}\ and\ \bibinfo {author} {\bibfnamefont {S.}~\bibnamefont {{Nesseris}}},\ }\href {\doibase 10.48550/arXiv.2412.01740} {\bibfield  {journal} {\bibinfo  {journal} {arXiv e-prints}\ ,\ \bibinfo {eid} {arXiv:2412.01740}} (\bibinfo {year} {2024})},\ \Eprint {http://arxiv.org/abs/2412.01740} {arXiv:2412.01740 [astro-ph.CO]} \BibitemShut {NoStop}%
\bibitem [{\citenamefont {Chevallier}\ and\ \citenamefont {Polarski}(2001)}]{Chevallier:2000qy}%
  \BibitemOpen
  \bibfield  {author} {\bibinfo {author} {\bibfnamefont {M.}~\bibnamefont {Chevallier}}\ and\ \bibinfo {author} {\bibfnamefont {D.}~\bibnamefont {Polarski}},\ }\href {\doibase 10.1142/S0218271801000822} {\bibfield  {journal} {\bibinfo  {journal} {Int. J. Mod. Phys. D}\ }\textbf {\bibinfo {volume} {10}},\ \bibinfo {pages} {213} (\bibinfo {year} {2001})},\ \Eprint {http://arxiv.org/abs/gr-qc/0009008} {arXiv:gr-qc/0009008} \BibitemShut {NoStop}%
\bibitem [{\citenamefont {Linder}(2003)}]{Linder:2002et}%
  \BibitemOpen
  \bibfield  {author} {\bibinfo {author} {\bibfnamefont {E.~V.}\ \bibnamefont {Linder}},\ }\href {\doibase 10.1103/PhysRevLett.90.091301} {\bibfield  {journal} {\bibinfo  {journal} {Phys. Rev. Lett.}\ }\textbf {\bibinfo {volume} {90}},\ \bibinfo {pages} {091301} (\bibinfo {year} {2003})},\ \Eprint {http://arxiv.org/abs/astro-ph/0208512} {arXiv:astro-ph/0208512} \BibitemShut {NoStop}%
\bibitem [{\citenamefont {Morawetz}\ \emph {et~al.}(2025)\citenamefont {Morawetz} \emph {et~al.}}]{DESI:2025hao}%
  \BibitemOpen
  \bibfield  {author} {\bibinfo {author} {\bibfnamefont {J.}~\bibnamefont {Morawetz}} \emph {et~al.} (\bibinfo {collaboration} {DESI}),\ }\href@noop {} {\  (\bibinfo {year} {2025})},\ \Eprint {http://arxiv.org/abs/2508.11811} {arXiv:2508.11811 [astro-ph.CO]} \BibitemShut {NoStop}%
\bibitem [{\citenamefont {\'O~Colg\'ain}\ \emph {et~al.}(2025)\citenamefont {\'O~Colg\'ain}, \citenamefont {Pourojaghi}, \citenamefont {Sheikh-Jabbari},\ and\ \citenamefont {Yin}}]{Colgain:2025nzf}%
  \BibitemOpen
  \bibfield  {author} {\bibinfo {author} {\bibfnamefont {E.}~\bibnamefont {\'O~Colg\'ain}}, \bibinfo {author} {\bibfnamefont {S.}~\bibnamefont {Pourojaghi}}, \bibinfo {author} {\bibfnamefont {M.~M.}\ \bibnamefont {Sheikh-Jabbari}}, \ and\ \bibinfo {author} {\bibfnamefont {L.}~\bibnamefont {Yin}},\ }\href@noop {} {\  (\bibinfo {year} {2025})},\ \Eprint {http://arxiv.org/abs/2504.04417} {arXiv:2504.04417 [astro-ph.CO]} \BibitemShut {NoStop}%
\bibitem [{\citenamefont {\'O~Colg\'ain}\ and\ \citenamefont {Sheikh-Jabbari}(2025)}]{Colgain:2024mtg}%
  \BibitemOpen
  \bibfield  {author} {\bibinfo {author} {\bibfnamefont {E.}~\bibnamefont {\'O~Colg\'ain}}\ and\ \bibinfo {author} {\bibfnamefont {M.~M.}\ \bibnamefont {Sheikh-Jabbari}},\ }\href {\doibase 10.1093/mnrasl/slaf042} {\bibfield  {journal} {\bibinfo  {journal} {Monthly Notices of the Royal Astronomical Society: Letters}\ }\textbf {\bibinfo {volume} {542}},\ \bibinfo {pages} {L24} (\bibinfo {year} {2025})},\ \Eprint {http://arxiv.org/abs/2412.12905} {arXiv:2412.12905 [astro-ph.CO]} \BibitemShut {NoStop}%
\bibitem [{\citenamefont {Sahni}\ \emph {et~al.}(2008)\citenamefont {Sahni}, \citenamefont {Shafieloo},\ and\ \citenamefont {Starobinsky}}]{Sahni:2008xx}%
  \BibitemOpen
  \bibfield  {author} {\bibinfo {author} {\bibfnamefont {V.}~\bibnamefont {Sahni}}, \bibinfo {author} {\bibfnamefont {A.}~\bibnamefont {Shafieloo}}, \ and\ \bibinfo {author} {\bibfnamefont {A.~A.}\ \bibnamefont {Starobinsky}},\ }\href {\doibase 10.1103/PhysRevD.78.103502} {\bibfield  {journal} {\bibinfo  {journal} {Phys. Rev. D}\ }\textbf {\bibinfo {volume} {78}},\ \bibinfo {pages} {103502} (\bibinfo {year} {2008})},\ \Eprint {http://arxiv.org/abs/0807.3548} {arXiv:0807.3548 [astro-ph]} \BibitemShut {NoStop}%
\bibitem [{\citenamefont {Foreman-Mackey}\ \emph {et~al.}(2013)\citenamefont {Foreman-Mackey}, \citenamefont {Hogg}, \citenamefont {Lang},\ and\ \citenamefont {Goodman}}]{Foreman-Mackey:2012any}%
  \BibitemOpen
  \bibfield  {author} {\bibinfo {author} {\bibfnamefont {D.}~\bibnamefont {Foreman-Mackey}}, \bibinfo {author} {\bibfnamefont {D.~W.}\ \bibnamefont {Hogg}}, \bibinfo {author} {\bibfnamefont {D.}~\bibnamefont {Lang}}, \ and\ \bibinfo {author} {\bibfnamefont {J.}~\bibnamefont {Goodman}},\ }\href {\doibase 10.1086/670067} {\bibfield  {journal} {\bibinfo  {journal} {Publ. Astron. Soc. Pac.}\ }\textbf {\bibinfo {volume} {125}},\ \bibinfo {pages} {306} (\bibinfo {year} {2013})},\ \Eprint {http://arxiv.org/abs/1202.3665} {arXiv:1202.3665 [astro-ph.IM]} \BibitemShut {NoStop}%
\bibitem [{\citenamefont {Lewis}(2019)}]{Lewis:2019xzd}%
  \BibitemOpen
  \bibfield  {author} {\bibinfo {author} {\bibfnamefont {A.}~\bibnamefont {Lewis}},\ }\href@noop {} {\  (\bibinfo {year} {2019})},\ \Eprint {http://arxiv.org/abs/1910.13970} {arXiv:1910.13970 [astro-ph.IM]} \BibitemShut {NoStop}%
\bibitem [{\citenamefont {{Wang}}(2024)}]{Wang:2024rjd}%
  \BibitemOpen
  \bibfield  {author} {\bibinfo {author} {\bibfnamefont {D.}~\bibnamefont {{Wang}}},\ }\href {\doibase 10.48550/arXiv.2404.13833} {\bibfield  {journal} {\bibinfo  {journal} {arXiv e-prints}\ ,\ \bibinfo {eid} {arXiv:2404.13833}} (\bibinfo {year} {2024})},\ \Eprint {http://arxiv.org/abs/2404.13833} {arXiv:2404.13833 [astro-ph.CO]} \BibitemShut {NoStop}%
\bibitem [{\citenamefont {Wang}\ and\ \citenamefont {Mota}(2025)}]{Wang:2025bkk}%
  \BibitemOpen
  \bibfield  {author} {\bibinfo {author} {\bibfnamefont {D.}~\bibnamefont {Wang}}\ and\ \bibinfo {author} {\bibfnamefont {D.}~\bibnamefont {Mota}},\ }\href@noop {} {\  (\bibinfo {year} {2025})},\ \Eprint {http://arxiv.org/abs/2504.15222} {arXiv:2504.15222 [astro-ph.CO]} \BibitemShut {NoStop}%
\bibitem [{\citenamefont {Wang}\ \emph {et~al.}(2025)\citenamefont {Wang}, \citenamefont {Yu},\ and\ \citenamefont {Wu}}]{Wang:2025vtw}%
  \BibitemOpen
  \bibfield  {author} {\bibinfo {author} {\bibfnamefont {J.}~\bibnamefont {Wang}}, \bibinfo {author} {\bibfnamefont {H.}~\bibnamefont {Yu}}, \ and\ \bibinfo {author} {\bibfnamefont {P.}~\bibnamefont {Wu}},\ }\href {\doibase 10.1140/epjc/s10052-025-14593-0} {\bibfield  {journal} {\bibinfo  {journal} {Eur. Phys. J. C}\ }\textbf {\bibinfo {volume} {85}},\ \bibinfo {pages} {853} (\bibinfo {year} {2025})},\ \Eprint {http://arxiv.org/abs/2507.22575} {arXiv:2507.22575 [astro-ph.CO]} \BibitemShut {NoStop}%
\bibitem [{\citenamefont {{\'O Colg{\'a}in}}\ \emph {et~al.}(2025)\citenamefont {{\'O Colg{\'a}in}}, \citenamefont {{Pourojaghi}},\ and\ \citenamefont {{Sheikh-Jabbari}}}]{Colgain:2024ksa}%
  \BibitemOpen
  \bibfield  {author} {\bibinfo {author} {\bibfnamefont {E.}~\bibnamefont {{\'O Colg{\'a}in}}}, \bibinfo {author} {\bibfnamefont {S.}~\bibnamefont {{Pourojaghi}}}, \ and\ \bibinfo {author} {\bibfnamefont {M.~M.}\ \bibnamefont {{Sheikh-Jabbari}}},\ }\href {\doibase 10.1140/epjc/s10052-025-13995-4} {\bibfield  {journal} {\bibinfo  {journal} {Eur. Phys. J \textbf{C}}\ }\textbf {\bibinfo {volume} {85}},\ \bibinfo {pages} {286} (\bibinfo {year} {2025})},\ \Eprint {http://arxiv.org/abs/2406.06389} {arXiv:2406.06389 [astro-ph.CO]} \BibitemShut {NoStop}%
\bibitem [{\citenamefont {Di~Valentino}\ \emph {et~al.}(2025)\citenamefont {Di~Valentino} \emph {et~al.}}]{DiValentino:2025sru}%
  \BibitemOpen
  \bibfield  {author} {\bibinfo {author} {\bibfnamefont {E.}~\bibnamefont {Di~Valentino}} \emph {et~al.},\ }\href@noop {} {\  (\bibinfo {year} {2025})},\ \Eprint {http://arxiv.org/abs/2504.01669} {arXiv:2504.01669 [astro-ph.CO]} \BibitemShut {NoStop}%
\bibitem [{\citenamefont {Akarsu}\ \emph {et~al.}(2024)\citenamefont {Akarsu}, \citenamefont {\'O~Colg\'ain}, \citenamefont {Sen},\ and\ \citenamefont {Sheikh-Jabbari}}]{Akarsu:2024qiq}%
  \BibitemOpen
  \bibfield  {author} {\bibinfo {author} {\bibfnamefont {O.}~\bibnamefont {Akarsu}}, \bibinfo {author} {\bibfnamefont {E.}~\bibnamefont {\'O~Colg\'ain}}, \bibinfo {author} {\bibfnamefont {A.~A.}\ \bibnamefont {Sen}}, \ and\ \bibinfo {author} {\bibfnamefont {M.~M.}\ \bibnamefont {Sheikh-Jabbari}},\ }\href {\doibase 10.3390/universe10080305} {\bibfield  {journal} {\bibinfo  {journal} {Universe}\ }\textbf {\bibinfo {volume} {10}},\ \bibinfo {pages} {305} (\bibinfo {year} {2024})},\ \Eprint {http://arxiv.org/abs/2402.04767} {arXiv:2402.04767 [astro-ph.CO]} \BibitemShut {NoStop}%
\bibitem [{\citenamefont {{Mukherjee}}\ and\ \citenamefont {{Sen}}(2024)}]{Mukherjee:2024pcg}%
  \BibitemOpen
  \bibfield  {author} {\bibinfo {author} {\bibfnamefont {P.}~\bibnamefont {{Mukherjee}}}\ and\ \bibinfo {author} {\bibfnamefont {A.~A.}\ \bibnamefont {{Sen}}},\ }\href {\doibase 10.48550/arXiv.2412.13973} {\bibfield  {journal} {\bibinfo  {journal} {arXiv e-prints}\ ,\ \bibinfo {eid} {arXiv:2412.13973}} (\bibinfo {year} {2024})},\ \Eprint {http://arxiv.org/abs/2412.13973} {arXiv:2412.13973 [astro-ph.CO]} \BibitemShut {NoStop}%
\bibitem [{\citenamefont {Teixeira}\ \emph {et~al.}(2025)\citenamefont {Teixeira}, \citenamefont {Giar\`e}, \citenamefont {Hogg}, \citenamefont {Montandon}, \citenamefont {Poudou},\ and\ \citenamefont {Poulin}}]{Teixeira:2025czm}%
  \BibitemOpen
  \bibfield  {author} {\bibinfo {author} {\bibfnamefont {E.~M.}\ \bibnamefont {Teixeira}}, \bibinfo {author} {\bibfnamefont {W.}~\bibnamefont {Giar\`e}}, \bibinfo {author} {\bibfnamefont {N.~B.}\ \bibnamefont {Hogg}}, \bibinfo {author} {\bibfnamefont {T.}~\bibnamefont {Montandon}}, \bibinfo {author} {\bibfnamefont {A.}~\bibnamefont {Poudou}}, \ and\ \bibinfo {author} {\bibfnamefont {V.}~\bibnamefont {Poulin}},\ }\href@noop {} {\  (\bibinfo {year} {2025})},\ \Eprint {http://arxiv.org/abs/2504.10464} {arXiv:2504.10464 [astro-ph.CO]} \BibitemShut {NoStop}%
\bibitem [{\citenamefont {Afroz}\ and\ \citenamefont {Mukherjee}(2025)}]{Afroz:2025iwo}%
  \BibitemOpen
  \bibfield  {author} {\bibinfo {author} {\bibfnamefont {S.}~\bibnamefont {Afroz}}\ and\ \bibinfo {author} {\bibfnamefont {S.}~\bibnamefont {Mukherjee}},\ }\href@noop {} {\  (\bibinfo {year} {2025})},\ \Eprint {http://arxiv.org/abs/2504.16868} {arXiv:2504.16868 [astro-ph.CO]} \BibitemShut {NoStop}%
\bibitem [{\citenamefont {G{\'o}mez-Valent}(2022)}]{Gomez-Valent:2022hkb}%
  \BibitemOpen
  \bibfield  {author} {\bibinfo {author} {\bibfnamefont {A.}~\bibnamefont {G{\'o}mez-Valent}},\ }\href {\doibase 10.1103/PhysRevD.106.063506} {\bibfield  {journal} {\bibinfo  {journal} {Phys. Rev. D}\ }\textbf {\bibinfo {volume} {106}},\ \bibinfo {pages} {063506} (\bibinfo {year} {2022})},\ \Eprint {http://arxiv.org/abs/2203.16285} {arXiv:2203.16285 [astro-ph.CO]} \BibitemShut {NoStop}%
\bibitem [{\citenamefont {\'O~Colg{\'a}in}\ \emph {et~al.}(2025{\natexlab{a}})\citenamefont {\'O~Colg{\'a}in}, \citenamefont {Pourojaghi}, \citenamefont {Sheikh-Jabbari},\ and\ \citenamefont {Sherwin}}]{Colgain:2023bge}%
  \BibitemOpen
  \bibfield  {author} {\bibinfo {author} {\bibfnamefont {E.}~\bibnamefont {\'O~Colg{\'a}in}}, \bibinfo {author} {\bibfnamefont {S.}~\bibnamefont {Pourojaghi}}, \bibinfo {author} {\bibfnamefont {M.~M.}\ \bibnamefont {Sheikh-Jabbari}}, \ and\ \bibinfo {author} {\bibfnamefont {D.}~\bibnamefont {Sherwin}},\ }\href {\doibase 10.1140/epjc/s10052-024-13727-0} {\bibfield  {journal} {\bibinfo  {journal} {Eur. Phys. J. C}\ }\textbf {\bibinfo {volume} {85}},\ \bibinfo {pages} {124} (\bibinfo {year} {2025}{\natexlab{a}})},\ \Eprint {http://arxiv.org/abs/2307.16349} {arXiv:2307.16349 [astro-ph.CO]} \BibitemShut {NoStop}%
\bibitem [{\citenamefont {Wilks}(1938)}]{Wilks:1938dza}%
  \BibitemOpen
  \bibfield  {author} {\bibinfo {author} {\bibfnamefont {S.~S.}\ \bibnamefont {Wilks}},\ }\href {\doibase 10.1214/aoms/1177732360} {\bibfield  {journal} {\bibinfo  {journal} {Annals Math. Statist.}\ }\textbf {\bibinfo {volume} {9}},\ \bibinfo {pages} {60} (\bibinfo {year} {1938})}\BibitemShut {NoStop}%
\bibitem [{\citenamefont {\'O~Colg{\'a}in}\ \emph {et~al.}(2025{\natexlab{b}})\citenamefont {\'O~Colg{\'a}in}, \citenamefont {Sheikh-Jabbari},\ and\ \citenamefont {Yin}}]{Colgain:2024clf}%
  \BibitemOpen
  \bibfield  {author} {\bibinfo {author} {\bibfnamefont {E.}~\bibnamefont {\'O~Colg{\'a}in}}, \bibinfo {author} {\bibfnamefont {M.~M.}\ \bibnamefont {Sheikh-Jabbari}}, \ and\ \bibinfo {author} {\bibfnamefont {L.}~\bibnamefont {Yin}},\ }\href {\doibase 10.1016/j.dark.2025.101975} {\bibfield  {journal} {\bibinfo  {journal} {Phys. Dark Univ.}\ }\textbf {\bibinfo {volume} {49}},\ \bibinfo {pages} {101975} (\bibinfo {year} {2025}{\natexlab{b}})},\ \Eprint {http://arxiv.org/abs/2405.19953} {arXiv:2405.19953 [astro-ph.CO]} \BibitemShut {NoStop}%
\bibitem [{\citenamefont {Trotta}(2017)}]{Trotta:2017wnx}%
  \BibitemOpen
  \bibfield  {author} {\bibinfo {author} {\bibfnamefont {R.}~\bibnamefont {Trotta}}\ }(\bibinfo {year} {2017})\ \Eprint {http://arxiv.org/abs/1701.01467} {arXiv:1701.01467 [astro-ph.CO]} \BibitemShut {NoStop}%
\end{thebibliography}%

\end{document}